\documentclass[12pt,letterpaper]{article}
\usepackage{amsmath}
\usepackage{amssymb}
\usepackage{latexsym}
\usepackage{hyperref}
\newcommand{\be}{\begin{equation}}
\newcommand{\ee}{\end{equation}}
\newcommand{\bea}{\begin{eqnarray}}
\newcommand{\eea}{\end{eqnarray}}

\begin{document} 

\title{Why Is There Something, Rather Than Nothing?\footnote{Invited contribution to the \textit{Routledge Companion to the Philosophy of Physics}, eds. E. Knox and A. Wilson (Routledge). CALT 2018-004.}}

 \author{Sean M. Carroll\\
 Walter Burke Institute for Theoretical Physics\\
 California Institute of Technology, Pasadena, CA 91125, U.S.A.\\
 seancarroll@gmail.com}
\maketitle


Science and philosophy are concerned with asking how things are, and why they are the way they are. It therefore seems natural to take the next step and ask why things \emph{are} at all -- why the universe exists, or why there is something rather than nothing \cite{holt,leslie}.

Ancient philosophers didn't focus too much on what Heidegger \cite{heidegger} called the ``fundamental question of metaphysics" and Gr\"unbaum \cite{grunbaum} has dubbed the ``Primordial Existential Question."
It was Leibniz, in the eighteenth century, who first explicitly asked ``Why is there something rather than nothing?" in the context of discussing his Principle of Sufficient Reason (``nothing is without a ground or reason why it is") \cite{leibniz}.
By way of an answer, Leibniz appealed to what has become a popular strategy: God is the reason the universe exists, but God's existence is its own reason, since God exists necessarily.
(There is a parallel with Aristotle's much earlier invocation of an unmoved mover, responsible for motion in the universe without itself being moved by anything else \cite{aristotle}.)

Subsequent thinkers were less impressed by this move. Hume \cite{hume1} explicitly dismissed the idea of a necessary being, and both he \cite{hume2} and Kant \cite{kant} doubted that the intellectual tools we have developed to understand the world of experience could sensibly be extended to an explanation for existence itself.
In their inimitable styles, Bertrand Russell \cite{russell2} shrugged off the question with ``I should say that the universe is just there, and that's all," while Ludwig Wittgenstein \cite{wittgenstein} suggested there were some things about which we should remain silent: ``It is not \emph{how} things are in the world that is mystical, but \emph{that} it exists."
More recently, Parfit \cite{parfit} argued for a middle ground between this kind of ``Brute Fact'' view and the idea that the universe is necessary, by suggesting that one or more features of our universe may pick it out as somehow special, even if they don't imply necessity.

The naturalness of the impulse to ask why the universe exists does not imply that the question is coherent or answerable. Reality is unique -- even if there are in some sense many existent ``worlds," we can take reality to be the single collection of all such worlds. It might be the case that the property of ``having a reason why" applies to facts within reality, but not to all of reality itself.
 
A major obstacle to addressing this question is the difficulty we have in putting aside strategies and assumptions that have served us well in less sweeping inquiries. Our experience of the world, which is confined to an extraordinarily tiny fraction of reality, doesn't leave us well equipped to think in appropriate ways about the question of its existence. On the contrary, it is very difficult to resist the temptation to treat the universe as just another thing, like an anteater or a smartphone, whose existence can be accounted for in relatively familiar ways. We should be constantly on guard not to insist on conventional answers for such a singular question.
 
Nevertheless, we can make some progress on the question of why reality exists, both by carefully considering what it might mean to obtain a convincing answer, and by looking at what modern physics and cosmology have taught us about the nature of the universe whose existence we are trying to explain. The most promising answer to date is that the existence of the universe is unlikely to be the kind of thing that has a reason why.

\section{What Does ``Why" Mean?}

Since at least the time of Aristotle, philosophers have developed elaborate taxonomies of different kinds of causes or explanations or reasons why various things are true. 
For our limited purposes here, it should suffice to distinguish between how the universe came to be and what \emph{mechanism} (if any) might have brought it into being -- corresponding roughly to Aristotle's efficient cause -- and the \emph{reason} why (if any) it exists -- corresponding to the final cause. 
Aristotle conceived of final causes teleologically, as ends or purposes. Here we're being a little broader, expanding the category to include anything that would qualify as a ``reason why." (See Section 8 of this volume for a discussion of a variety of approaches to explanation in physics, and \cite{skow} for a recent philosophical investigation of reasons why.) Let us label these notions ``mechanisms'' and ``reasons'' for short. The categories are not exclusive. For some purposes, pointing to a mechanism might be enough to answer a question about why something happened; for others, a deeper kind of reason might be sought.

``What mechanism brought the universe into existence?" is conceivably a scientific question. Keeping in mind that ``there is no such thing" is a perfectly plausible answer, attempts to identify a mechanism that brought the universe into being should, at the very least, be informed by our best current science, and ideas from contemporary physics have significantly affected what kind of answer we might reasonably expect.

``What reason explains why anything exists at all?" is another matter entirely. Aristotle treated final causes as a fundamental metaphysical category, an irreducible feature of the architecture of reality. Modern physics sees things differently. Rather than being a story of effects and their associated causes, the universe is described by patterns, called the laws of physics, that relate conditions at different times and places to each other, typically by differential equations. (See Lange [this volume] for more discussion of the nature of laws.) The difference between the two conceptions is that the former naturally associates things that happen with a deeper kind of reason why they do, while on the latter view every ``why" question is definitively answered by ``the dynamical laws of nature and the initial conditions of the universe." The idea that laws simply describe patterns, rather than actively governing what is allowed, is known in the philosophy literature as a ``Humean account" of the laws of nature \cite{hall}. 

Scientists are still happy, of course, to talk about explanations and reasons why, in at least two contexts: when accounting for some particular state of affairs in the context of a higher-level (emergent) description, and when pointing to underlying principles as providing explanations for properties of the universe or its dynamics.
Fundamental physics explains states of affairs by reference to the initial conditions of the universe, but in emergent theories involving coarse-graining of microscopic degrees of freedom, it can be natural to point to specific effects as arising from individual causes \cite{russell,norton,hitchcock,carroll}.\footnote{The emergent nature of causality can be traced in part to the fact that the entropy of the universe was very low in the past, which gives us great leverage over associating past ``causes'' with present ``effects,'' in a way that isn't available for future events \cite{albert,loewer}. Fundamental physics often deals with small numbers of degrees of freedom, where entropic considerations aren't relevant and there is no arrow of time. See also Shahvisi (this volume).} 
When it comes to properties of the world rather than states of affairs, explanations often take the form of appeal to symmetries or other deeper principles: we say that conservation of charge is explained by the symmetry of gauge invariance in electromagnetism.

What does this mean for the existence of the universe? 
If cause-and-effect language as applied to states of affairs can usefully emerge at higher levels but is absent in fundamental physics, looking for the ``cause'' of the universe would be a pointless endeavor.
By construction, the universe is the most fundamental thing there is.\footnote{I'm using ``universe'' here to refer to the entirety of physical reality. No judgment is implied about whether things other than physical reality can be usefully said to exist. This definition of universe would include all branches of the wave function in Everettian quantum mechanics, and all parts of a cosmological multiverse. For more discussion of fundamentality, see (French, this volume).} 
The best we can ask is whether we can imagine laws of nature that fully account for how the universe behaves, even at the earliest moments, or whether we are forced to look outside of reality itself in search of some kind of cause.
While we don't currently know the once-and-for-all laws of nature, nothing that we do currently understand about physics implies any necessary obstacle to thinking of the universe as a fully law-abiding, self-contained system.
In this case, there would be no such thing as the ``cause'' or ``reason why'' the universe exists, even if such notions are appropriate when talking about why a glass falls to the floor or why do fools fall in love.
The latter examples are embedded within larger explanatory contexts, while reality is not.

In the second sense of explanation, accounting for properties of nature by appeal to deeper principles, we might hope to find purchase.
That is, there might be something special about the way our universe is, which we could then point to as the reason why it exists.
Perhaps it is the minimal imaginable universe, or the most symmetric, or the most elegant, or even the only possible universe (presumably subject to some reasonable conditions).
If some such principle were to be found, we would still have to worry that the question was simply kicked up a level: \emph{why} does the universe satisfy this particular criterion?
Demonstrating that reality was the simplest or most beautiful example among a certain class of possible realities might gently scratch some explanatory itch, but we would be left wondering why such a principle should be given credit for bringing the universe into existence at all. 
Why shouldn't the universe be ugly or baroque?
It would be different if our universe were the only possible one, but as we will see that possibility has it's own problems.

This kind of worry generalizes into a concern about \emph{explanatory regression}: given any purported reason why reality exists, why is that reason valid?
One option, following Leibniz and others, is that we reach a level at which further explanation is not required, because something is necessarily true.
At the other end of the spectrum, explanations might bottom out with a brute fact: something that simply is the case, without further reason, even though it didn't necessarily have to be that way.
Arguably there is an in-between stance, where there is something that isn't strictly necessary, but nevertheless satisfies some principle (perhaps of simplicity or beauty) that qualifies as at least a partial explanation.
We should be aware of all of these possibilities while examining how our universe might ultimately be explained.

Given these considerations, there is a list of options that might conceivably qualify as an answer to ``Why is there something rather than nothing?":
\begin{itemize}
\item \textit{Creation:} There is something apart from physical reality, which brings it into existence and/or sustains it. This hypothetical entity is often identified with God in the literature, but there is not necessarily any strong connection with a traditional theistic conception of the divine. 
\item \textit{Metaverse:} Just as we can sometimes explain events within the universe by appeal to a causal web describing the universe as a whole, perhaps what we think of as reality is part of a larger context, a metaverse that could help explain the existence and properties of our universe. (We're imagining here something more profound than the traditional cosmological multiverse, which is just a universe in which conditions are very different in different regions of spacetime.)
\item \textit{Principle:} There is something special about reality, in that it satisfies some underlying principle, perhaps of simplicity or beauty. 
\item \textit{Coherence:} Perhaps the concept of ``nothingness'' is incoherent, and the possibility of reality not existing was never actually a viable option. 
\item \textit{Brute fact:} Reality itself simply exists, in the way that it does, without further explanation.
\end{itemize}
We can keep these alternatives in mind as we consider further background issues, before coming back to evaluating them at the end.

\section{What Do ``Something'' and ``Nothing" Mean?}

One place where science has exerted an impact on the question is in our definitions of ``something" and ``nothing."
In olden times, we might have described the universe as a collection of stuff (matter, energy, fields), distributed through space and evolving with time.
We can then distinguish between two issues:
\begin{enumerate}
\item Why is there stuff? Why is there anything inside the universe, rather than just empty space?
\item Why is there space at all? Why is there anything we would recognize as ``a universe"?
\end{enumerate}
For the first question, the relevant notion of ``nothing" is ``empty space," while for the second it is the non-existence of reality altogether.
Clearly it's the second question that most people have in mind when they ask why there is something rather than nothing, but answers to the first question (which are much easier to imagine obtaining) have often been passed off as relevant to the second.

Newtonian mechanics provides a precise mathematical formalization of this picture. (See [Samaroo, this volume] for more details.)
In the absence of external intervention, Newtonian absolute space is eternal, since the equations of motion can be extended infinitely far into the past or future.
There is no natural context in which to talk about the creation of the universe, without explicitly invoking divine intervention or something equivalent.
(Newton himself thought of God as creating the universe and sustaining its existence, even occasionally intervening when appropriate, but the need for anything outside the universe was rejected by Laplace and other subsequent Newtonian thinkers.) 

With the advent of special relativity, space and time are combined into spacetime, and in general relativity spacetime becomes dynamical and responsive to the presence of matter and energy.
The basic paradigm remains the same, with one important exception: spacetime itself can begin or end, in a Big Bang or Big Crunch singularity, and indeed the simplest models of our observed universe suggest that there was such a singularity in the past.
General relativity therefore diverges from Newtonian absolute space and time in allowing for a universe with a beginning, a first moment in time.
It is tempting to think of this as a transition from nothing to something; it is also tempting to think that a universe that begins calls out for an explicit cause more than an eternal universe would -- otherwise why did it begin?
We'll talk more about the wisdom of giving into these temptations in the next section.

The bigger shift came with the introduction of quantum mechanics.
Here we face the problem that there is no consensus about what the ultimate ontology of quantum mechanics actually is; there are various competing ``interpretations,'' which really amount to distinct physical theories.
Common to them all is the idea of a wave function of a system, which provides us with the probability of obtaining specified measurement outcomes.
In an Everettian or Many-Worlds approach, that wave function is all there is, and it splits into branches describing effectively separate worlds when subsystems become entangled and decohere. (See \cite{Wallace:2012zla} and [Saunders, this volume].)
In a hidden-variables approach such as the de~Broglie-Bohm theory, we posit additional degrees of freedom whose evolution is simply guided by the wave function. (See \cite{bohm} and [Tumulka, this volume].)
Here I will focus on the wave-function-only ontology.

Non-relativistic quantum mechanics isn't that different from Newtonian physics, as far as something vs.\ nothing is concerned.
There are a fixed number of particles, and the wave function describes the probability of observing particular values of their positions or momenta or other variables.
Relativity requires that we move to quantum field theory, which is a particular version of quantum mechanics in which the classical variables that are quantized to give a wave function are a set of field values throughout space, rather than positions or momenta of individual particles (see Section 5 of this volume).
The allowed states of the theory include a ``vacuum,'' defined as the lowest-energy state, and excited states describing collections of particles.
But the notion of the vacuum is subtle, as ``empty space'' isn't quite the same as ``nothing there."
Even in the emptiest lowest-energy state, there are still field degrees of freedom at every point in space, in a particular quantum configuration.
These degrees of freedom are highly entangled with each other, and can be probed by measurement devices.
For example, the Unruh effect describes the phenomenon by which an accelerated observer in the vacuum will detect a thermal bath of particles \cite{unruh}. 
Even more impressively, the Reeh-Schleider theorem establishes that any global quantum state of the system as a whole can be reached (to arbitrary precision) by starting with the vacuum and acting with some operator confined to a small region of space \cite{reeh}.
In other words, because field degrees of freedom in different regions of space are entangled in the vacuum, operating on the ones in any particular region can effectively produce any possible state of the theory.

There can also be multiple kinds of vacua in a single quantum field theory: a true vacuum that is the lowest-energy state, and false vacua that have no particles in them, but whose energy density is higher than in the true vacuum. (See also Castellani and Dardashti [this volume].)
Due to the phenomenon of spontaneous symmetry breaking \cite{ssb}, the most symmetric vacuum (in which the expectation value of all the quantum fields vanishes) is generally not the true vacuum.
Cosmological evolution plausibly involves a transition from a symmetric vacuum state, free of particles, to a collection of particles in a background given by a lower-energy vacuum. 
In some models, this evolution could dynamically favor matter over antimatter, helping to explain the current asymmetry in our observed universe.
Such a scenario has given rise to the pithy saying that there is something rather than nothing because ``nothing is unstable" \cite{wilczek,krauss}, if we allow ourselves the freedom to define ``nothing'' as ``a symmetric false-vacuum state."
This has nothing at all to do with the origin of the universe itself, and certainly nothing to do with why there is a quantum wave function in the first place.

In the context of creation of something from nothing, we must also face the issue of ``quantum fluctuations." 
(For a discussion of the very different senses in which this term can be used, see \cite{Boddy:2015fqa}.)
It is often said that the quantum vacuum is filled with fluctuating virtual particles, and even that these particles sometimes pop into real existence, as in Hawking radiation from black holes \cite{Page:2004xp}.
This is a misleading description, arising from a tendency to speak as if wave functions represent statistical ensembles of classical particles, rather than true quantum states.
A quantum state is simply a quantum state, and a true vacuum state will be stationary, with nothing ``fluctuating'' at all.
Hawking particles can be emitted by black holes because a state with a black hole is not the vacuum state, and the wave function of a black hole state naturally evolves into one with particles radiating away as the black hole shrinks.

The situation diverges from our Newtonian intuition even more dramatically when we turn to quantum gravity, in which spacetime itself has a wave function (see section 6 of this volume).
In that case there is no single ``spacetime,'' there are only approximate notions of spacetime that apply in a classical limit.
We do not yet have a full comprehensive theory of quantum gravity, but we can nevertheless try to make progress on the basis of general properties of gravity and quantum mechanics individually.
One consequence of quantum gravity is that the distinction between ``empty space" and ``space filled with stuff" is blurred, practically to invisibility.
An intriguing modern idea is that spacetime itself can be defined in terms of the entanglement between a set of abstract quantum degrees of freedom \cite{VanRaamsdonk:2010pw,Maldacena:2013xja,ccm}.

The other relevant consequence of quantum gravity is for the beginning of the universe.
Classically, there is good reason to believe that general relativity breaks down at a Big Bang singularity, which provides a boundary to spacetime in the past.
(The Big Bang isn't a point in space, but should be thought of as a spacelike surface; one is tempted to say ``a moment in time,'' except that spacetime is singular so ``time'' is not well-defined.)
But in a world with quantum mechanics, the breakdown of a classical theory simply means that we shouldn't be taking the classical limit as an accurate description of the situation at that point.\footnote{A theorem by Borde, Guth, and Vilenkin \cite{Borde:2001nh} demonstrates that spacetimes with an average expansion rate greater than zero must be geodesically incomplete in the past.
This is sometimes offered as an argument that the universe had a beginning, but that is incorrect.
Trivially, the average expansion rate could be zero, as it would be in a bouncing cosmology.
More importantly, the theorem only applies to classical spacetimes, so at most it could indicate where the classical approximation breaks down, not where the universe begins.}

The best we can say is that our current incomplete understanding of quantum gravity is fully compatible with both the possibility that the universe has lasted forever, and that it had a first moment in time.
We will look at both possibilities more carefully in the next section.
To understand why there is something rather than nothing, we certainly have to understand why there is a physical world described by a quantum wave function at all, and we might possibly have to understand how such a universe could ``come into existence'' out of nothing.

\section{The Possibility Question: Can the Universe Simply Be?}

We can now turn to the question proper: why is there something rather than nothing?
The first issue to be addressed is whether physical reality \emph{requires} something external to itself to account for its existence: either something to sustain it, if the universe exists eternally, or something to bring it into existence, if the universe had a beginning.
We can consider each scenario in turn.

For definiteness let's imagine that some form of quantum mechanics is the correct description of the physical world at its most fundamental level. That might not be true, but arguably the lessons we learn will generalize to other ontologies.
A quantum state $|\Psi\rangle$ is a vector in a Hilbert space $\mathcal{H}$. 
(A Hilbert space is essentially just a vector space with an inner product defined between vectors.)
We posit a Hamiltonian operator $\widehat{H}$ that defines the energy of a state, and then the dynamics of the theory are described by Schr\"odinger's equation
\be
  \widehat{H}|\Psi\rangle = i\hbar \frac{\partial}{\partial t}|\Psi\rangle.
  \label{se}
\ee
This equation applies to the dynamics of any isolated quantum system, including relativistic quantum field theories and presumably quantum gravity; all one has to do is specify the right Hilbert space and Hamiltonian.
(We assume the universe is isolated, or else we should be including whatever influences it as part of the universe.)

The Schr\"odinger equation has an immediate, profound consequence: almost all quantum states evolve eternally toward both the past and the future.
Unlike classical models such as spacetime in general relativity, which can hit singularities beyond which evolution cannot be extended, quantum evolution is very simple.
Any state can be written as a superposition of states of definite energy (eigenstates), in terms of which the Schr\"odinger equation implies that the magnitude of each coefficient remains constant, while the phase orbits at a fixed velocity (at least for time-independent Hamiltonians).
In Hilbert space, the entire evolution of the universe simply describes eternal motion in a straight line within some high-dimensional space that is topologically a torus \cite{Carroll:2008yd}.

If this setup describes the real world, there is no beginning nor end to time.
This is not to say that there is no Big Bang in the usual sense; only that it is not a true physical singularity as it would be in classical general relativity, nor does it represent the first moment of the universe.
As far as physics is concerned, such a universe would be completely self-contained, existing perpetually without any external cause.
One can still question whether or not an uncaused eternal universe is intellectually satisfying, but there is no physical or cosmological obstacle to its existence.

This situation applies to ``almost all'' quantum states, because there is an exception: states with exactly zero energy.
Then Schr\"odinger's equation collapses to
\be
  \widehat{H}|\Psi\rangle = 0,
  \label{wdw}
\ee
which in the context of quantum gravity is known as the Wheeler-DeWitt equation \cite{wdw}.
An equation of this form arises directly from a straightforward attempt to apply the usual rules of canonical quantization to general relativity.
There is nothing special here about the quantization procedure; the Wheeler-DeWitt equation simply reflects the fact that general relativity is invariant under reparameterizations of the time coordinate, a feature which exists even in the classical theory \cite{Rovelli:2015gwa}.
We should not imagine that we have any firm reason to expect that an equation of this form \emph{must} be the foundational relation of quantum gravity; it is certainly plausible that the more general form (\ref{se}) governs the evolution of the quantum state of the universe, and that the symmetries of general relativity are approximate and emergent in the classical limit.

If the Wheeler-DeWitt equation (\ref{wdw}) is correct, it presents us with an immediate challenge, known as the ``problem of time": there is no time parameter in the equation, so what is ``time" supposed to mean? (See [Th\'ebault, this volume] for more details.)
One might think that such an equation is ruled out by experiment, since we experience the passage of time in the real world.
But time might be emergent, rather than fundamental.
(Emergent time can still be ``real,'' just as a fluid is still real even if it emerges out of the collective behavior of atoms. See [Huggett, this volume] for more discussion.)
In other words, we can imagine factorizing Hilbert space into a tensor product of the form
\be
  \mathcal{H} = \mathcal{H}_C \otimes \mathcal{H}_U,
\ee
where $\mathcal{H}_C$ describes some ``clock'' subsystem of the universe, and $\mathcal{H}_U$ describes everything else.
Then it's possible to define variables such that the state restricted to $\mathcal{H}_U$ is entangled with the clock, so that the rest of the universe seems to ``evolve" according to some emergent equations of motion \cite{Page:1983uc,Banks:1984cw,ib}.
This procedure faces a somewhat under-appreciated problem, however, known as the ``clock ambiguity": one can factorize Hilbert space in many different ways, obtaining different kinds of emergent time evolution \cite{Albrecht:2008bj}.
It is far from clear how this ambiguity should be resolved, and consequently unclear whether the Wheeler-DeWitt equation by itself can serve as the basis for a well-formed theory of quantum gravity.

Because time is emergent in such models, it might only stretch over a finite interval, so the universe might not be eternal (though in some models it still could be).
The Wheeler-DeWitt equation has therefore been used as the basis for models in which the universe has an earliest moment of time \cite{Vilenkin:1983xq,Hartle:1983ai}.
Sometimes, such universes are said to ``come into existence out of nothing."
This is a misleading way of putting it, as it implies a temporal process that begins with nothing and ends with the universe.
But if the universe doesn't exist, there is no time, and hence there are no processes.
It is better, instead, to reserve temporal vocabulary for that portion of reality over which time actually exists.
The question is not whether a universe could pop into existence out of nothingness, but whether a universe with a beginning can be entirely described by an appropriate set of laws of physics without the help of any external cause.
The answer is that, by itself, the existence of an earliest moment to time is no obstacle to describing the physical universe in completely consistent, self-contained terms.
There is therefore no requirement, at least as far as physics is concerned, that existence have an identifiable cause independent of physical reality, whether the universe stretches infinitely far back in time or only a finite interval.

This exposition has been somewhat technical, but we can construct a more intuitive explanation based on the concept of conservation of energy.
Energy conservation can be thought of as the idea that the energy of a system at one moment is exactly the same as it was at a moment immediately before, and will be at a moment immediately after.
If the energy is nonzero, therefore, it follows that time must extend in both directions -- the energy must ``go somewhere'' in time.
But the Wheeler-DeWitt equation describes a universe whose total energy is exactly zero, one where gravitational energy is precisely equal in magnitude and opposite in sign to the energy of matter and other sources. (This does not represent a delicate fine-tuning; it is automatically true in a universe where space is closed, for example a three-dimensional sphere.)
Such universes are precisely the kind where time need not flow forever -- effectively, a non-existent universe has the same energy (zero) as an existent one obeying the Wheeler-DeWitt equation -- and which can therefore have a beginning.
Whether or not we live in such a universe is still an open question.

These scientific considerations could be countered by an insistence that differential equations might \emph{describe} what the universe does, but they don't explain the \emph{reason why} it does those things.
That is true as far as it goes (why these equations, rather than some other ones? why equations at all?), but it is sometimes extended to a demand that such an explanation \emph{must} exist.
Demands of this sort often refer to Leibniz's Principle of Sufficient Reason (PSR) or a modern version thereof: everything must have a reason or explanation, including the universe itself. 
To avoid an infinite regress, one can suggest that while the universe itself is \emph{contingent} (it did not have to exist in its own right), the ultimate explanation for it can be found in a \emph{necessary} being \cite{necessary}.
Necessary beings, so the idea goes, don't themselves require any further explanations or causes.

From a modern perspective, arguments of this sort are not very convincing, as the justification for the PSR is somewhat antiquated. (Leibniz's own justification relied heavily on his view of God, and there remains a correlation between acceptance of the PSR and theism, though it is not a necessary connection.)
Once we think of the laws of nature as describing patterns rather than causal forces, and the notion of cause and effect as being appropriate to higher-level emergent descriptions of the world rather than the fundamental level, the PSR loses its luster.
It is sometimes defended as a prerequisite for understanding and talking about the universe at all: if things happen without reasons, how can we possibly make any sense of the world?\footnote{Cosmologists do sometimes imagine universes in which typical living creatures are ``Boltzmann Brains,'' random fluctuations out of the surrounding high-entropy chaos, and such models may indeed be rejected on the basis of cognitive instability \cite{bb}.}
But the requirement that the world be orderly and intelligible is much weaker than the demand that everything has a cause or reason behind it; there is a sizable gap between the PSR as usually understood and ``anything goes.''
In particular, somewhere in between is the idea of an orderly universe which follows impersonal, unbreakable patterns -- precisely the kind of universe that is described by modern physics.
Such a property is more than enough to allow for sensible investigation and discussion of how the world is, without implying the existence of anything outside the world; as we've seen, there is no shortage of ways the physical world could be both orderly and self-contained.

The idea of a necessary being is similarly unconvincing; there is a vast theological literature on the subject, and I will only offer the barest discussion here.
One route to arguing in its favor would be a ``cosmological argument," as briefly given above: everything requires a reason, and a necessary being would ultimately ground such reasons.
If the PSR (or an equivalent assumption) isn't fundamental, this argument is deprived of its force, as the existence of the universe wouldn't require a reason in the conventional sense.
An alternative is an ``ontological argument," pioneered by Anselm of Canterbury and offered up in modified versions frequently since.
The basic idea is to argue that we can conceive of a most perfect being, and to exist is more perfect than to not exist, therefore the most perfect being necessarily exists \cite{ontological}.
Of the several different objections one might raise, a strong one is to point out that we actually \emph{can't} conceive of a most perfect being, as the notion of ``perfection'' is not rigorously defined.
It has long been recognized that ontological arguments rarely convince skeptics of the need for a necessary being; as Alvin Plantinga admits about his own proof, it serves as a demonstration of the logical consistency of belief in God, not a requirement for doing so \cite{plantinga}.
The skeptics seem to be on firm ground; as Hume emphasized, there is no being whose non-existence would entail a logical contradiction, and we have no difficulty in conceiving of worlds in which no such being existed.

The idea of a universe created by a greater being, for some specific purpose or having some particular properties, seems somehow more \emph{satisfying} than a universe that existed without a brute fact.
(Our idea of satisfying explanations has, needless to say, been trained on our experience within a tiny fraction of reality, not on the existence of the whole of reality itself; but we work with what we have.)
Moreover, the presence of regularities such as the laws of nature is itself something we might want to explain, even if it alone is sufficient to render the universe intelligible. 
We are therefore welcome to search for evidence for such an extra-universal entity, using the conventional methods of science and reason.
But there is no logical or empirical reason why such an entity must exist; the universe can just be.

\section{The Naturalness Question: Why This Particular Universe?}

Even if the universe can simply be, we can ask why it exists in this way -- what explains the specific properties of the laws of nature and the arrangement of stuff in the cosmos.
This has been a longstanding goal of scientists; as Einstein \cite{einstein} famously put it, ``What really interests me is whether God had any choice in the creation of the world."
This phrasing points to an even farther-reaching ambition: to not only reveal the reason why the universe is this particular way, but potentially to discover that this way is unique, that there is literally no other way the universe could have been.
Whether we classified such a prospective discovery as providing the reason why the universe exists, or removing the need for such a reason, it would certainly satisfy the goal of understanding its existence.

Taken at face value, this ambition seems hopeless.
There are an infinite number of self-consistent quantum-mechanical systems that are different from our actual universe. (In terms of Schr\"odinger's equation, these different theories correspond to different sets of eigenvalues for the Hamiltonian operator.)
And there are presumably an infinite number of ways the laws of physics could have been that aren't quantum-mechanical at all.\footnote{I am glossing over a distinction one might draw between ``conceivable" and ``possible." The latter could be a narrow category than the former, if we added extra requirements onto the condition of possibility, such as logical consistency or agreement with the laws of nature. Neither of these criteria is relevant here (any Hamiltonian leads to logically consistent laws of physics, and the laws of nature are precisely what are being decided upon).}
A more sensible hypothesis might be that the universe and its laws of nature are the simplest that they could be, given that they also satisfy some other condition -- something as specific as describing a quantum-mechanical four-dimensional spacetime with local laws of physics, or something as broad as the existence of intelligent observers.

It is interesting to speculate whether the laws of physics governing reality as we experience it are in some (to be specified) sense maximally elegant, at least given some basic requirements.
The general trend of scientific discovery over the last few centuries has been to explain disparate complex phenomena in terms of comparatively simple and powerful frameworks.
Newtonian mechanics unified a wide range of phenomena in classical physics; Maxwell's electromagnetism provided a single explanation for light, radiative heat, electricity, and magnetism; Darwinian evolution brought diverse species under the umbrella of a single history of life on Earth; Einstein's relativity and modern quantum field theory used the power of symmetries to provide a simple account of numerous features of the laws of physics; and today we know that the wide variety of phenomena in our everyday lives can be thought of as different manifestations of just a few elementary particles interacting through a handful of forces.
Perhaps this trend can continue to an ultimate point where we find that all of the laws of physics applying to our universe can be encapsulated in a single succinct principle.

It's an attractive prospect, which may or may not be true.
Many observed features of both fundamental physics and cosmology seem to be arbitrary, from the large-scale structure of stars and galaxies to the masses of elementary particles.
Many physicists now suspect that the laws of physics in our observable universe are just one possibility among a very large ``landscape''  of physically realizable possibilities, known as ``vacua'' (since they are local minimum-energy states), each of which features different particles, forces, couplings, and even numbers of spatial dimensions \cite{Susskind:2003kw}.\footnote{There is the separate issue of whether such possibilities are actually realized in the form of a multiverse. This is a model-dependent question; within inflationary cosmology it doesn't seem difficult to imagine that all or most of the vacua are realized \cite{Linde:1986fd,Guth:2000ka,Hawking:2006ur}, though the issue isn't settled \cite{Banks:2003es,Johnson:2008vn}. There may also be a relationship between the cosmological multiverse and the many worlds of Everettian quantum theory \cite{Bousso:2011up,Nomura:2012zb}.}
In string theory, estimates for the size of this landscape throw around numbers of the form $10^{500}$.
One might take the attitude that the underlying equations of string theory are somehow maximally elegant, even if the specific low-energy manifestation of them that we observe is not.
But the lesson is that at present the idea that the ultimate laws are as simple as possible is a hope, not something suggested by the evidence.
Moreover, the prospect still faces the challenge of explanatory regression, as one would left to explain why the underlying laws should be so simple.

Another strategy is to point to the existence of intelligent life in the universe.
The ``anthropic principle'' \cite{barrow,Hogan:1999} is the idea that certain features of our observable environment are best explained by realizing that things had to be that way in order to allow for the existence of intelligent life.
There are various versions of this kind of reasoning, the most respectable of which uses anthropics as a kind of environmental selection: given an ensemble of many different kinds of conditions (regions of space, branches of the quantum wave function, or truly separate universes), we are guaranteed to find ourselves in the subset of those conditions that allow for intelligent life.
Could we explain the reason why we live in this universe, rather than some other kind of universe, on purely anthropic grounds?
(See Barnes [this volume] for more discussion of these strategies.)

There is a substantial obstacle here, over and above the evident difficulty in understanding what conditions actually  allow for the existence of life.
First, given some parameter such as the mass of a particular particle or the average energy density of the universe, there will be a range of values that are anthropically acceptable.
In the absence of any other considerations, we would predict that the parameter in question should look ``typical" within that range.
For example, the vacuum energy (or cosmological constant) is much smaller in magnitude than the naive value it might have, near the Planck scale of quantum gravity \cite{Carroll:2000fy}.
But if it were large and positive, the vacuum energy would cause such rapid acceleration of space that galaxies couldn't form, making it hard for life to exist; if it were large and negative, the universe would recollapse so rapidly that there would be no time for life to evolve.
Such reasoning was used by Weinberg to predict, against the expectations of many theoretical physicists, that astronomers would eventually detect a nonzero value for the vacuum energy \cite{Weinberg:1987dv}.
A decade later, that's exactly what they did, with the observed value seeming roughly typical within the allowed range \cite{Riess:1998cb,Perlmutter:1998np}.

For other parameters, however, this anthropic expectation predicts something very different from the real universe.
An obvious example is the low entropy of the early universe \cite{Penrose:1980ge,whatsense}, which is many orders of magnitude smaller than what it would need to be in order for life to exist.
More generally, the universe simply seems to have far more stuff in it than any reasonable anthropic criterion would imply; there are more than a trillion galaxies, with of order a hundred billion stars and planets in each of them, none of which is necessary for our existence here.
If the universe were minimal subject to the existence of intelligent life, why wouldn't it take the form of a relatively small collection of atoms in otherwise empty space, enough to come together to form a small number of stars and planets, before eventually dispersing back into the void?

The fact that our universe doesn't look as minimal as it possibly could, even conditioning on the existence of life, is a strike against one potentially promising answer to the question of why there is something rather than nothing: that every possible world actually exists, and ours is simply the one in which we happen to find ourselves \cite{lewis,tegmark}.
It's hard to know precisely what the set of all possible worlds looks like, and even harder to imagine putting a measure on it from which one could extract probabilisitic anthropic predictions.
Nevertheless, it seems reasonable that most intelligent observers would find themselves in worlds that were much less profligate with matter and energy than ours is, in a version of the Boltzmann Brain problem.
The safest tentative conclusion to draw is that the properties of our particular universe cannot be solely attributed to the fact that intelligent observers exist within it, even if some particular properties may be.

\section{The Reason Question: Why Does Anything Exist at All?}

Having laid this groundwork, we can at last turn to the question of why anything exists at all.
Let's consider the five options previously mentioned -- creation, metaverse, principle, coherence, and brute fact -- and briefly evaluate each of them.

\begin{itemize}
\item \textit{Creation.} 
\end{itemize}

The idea that our reality was brought into existence by some being outside of reality is perhaps the most intuitively appealing explanation for its existence.
For one thing, even if the universe could exist as a brute fact, existence is arguably not what we would expect; as Swinburne \cite{swinburne} has put it, ``It is extraordinary that there should exist anything at all. Surely the most natural state of affairs is simply nothing."

``Natural'' presumably isn't the right idea in this context; by definition, whatever reality is, it's natural.
What is meant is probably something the the effect that non-existence is simpler or easier than existence, and for some reason is therefore to be expected.
In part this expectation comes from our experience within the universe, where things typically need to be created and perhaps maintained.
Consciously or not, we have in mind a metaphorical reality-chooser, who contemplates all the different ways the universe could have been (perhaps including non-existent) and makes a simple and elegant choice.
Similarly, it is sometimes suggested that the regularities we label ``laws of nature" are inexplicable in the absence of some entity that ensures those laws are obeyed, as if the reliability of such laws implies the existence of a legislative body or a law-enforcement agency.
It is precisely this kind of intuitive and metaphorical reasoning that we should be suspicious of in this context. 
Ideas that become ingrained from our experience with the everyday world may not extend in any useful way to the very unique question of the existence of reality.

While a creator could explain the existence of our universe, we are left to explain the existence of a creator.
In order to avoid explanatory regression, it is tempting to say that the creator explains its own existence, but then we can ask why the universe couldn't have done the same thing.\footnote{All else being equal, a self-explaining and necessary universe would be a simpler overall package than a self-explaining and necessary creator who then created the universe. But to most advocates of this general strategy, necessity seems like a more natural property to attribute to a supernatural creator than to the natural universe.}
Thus we are left to identify the creator as a necessary being, in contrast with the contingent nature of our universe.
But as we argued at the end of Section 3, the idea of a necessary being doesn't really hold together; there just isn't any such thing.

The conclusion is that invoking a creator does not provide us any escape from the need to posit something that simply exists because it does, without further reasons to which we can appeal.
And if that is the case, there is no reason not to include all of reality in that category, without additionally imagining a creator at all.
The existence of a creator of the universe should be judged on ordinary empirical grounds (does it provide a useful explanatory account of observed features of what we see?), not on \emph{a priori} arguments for its necessity.

\begin{itemize}
\item \textit{Metaverse.} 
\end{itemize}

Cosmologists use the word ``multiverse'' to refer to something that is actually more prosaic than it sounds: a single connected spacetime, but with regions (``universes") where conditions are very different from each other.
In other contexts, including philosophical writing as well as the popular media, the word is sometimes applied to the multiply-branched wave function of Everettian quantum mechanics.
We have in mind something a bit more grandiose: a collection of truly distinct realities (noninteracting, not stemming from a common past, not necessarily with the same laws of physics), one of which is our own. 
We therefore label such a collection the ``metaverse."

The cosmological multiverse can provide a context in which anthropic reasoning becomes appropriate, and may therefore help explain why our observed region of space has some of the properties that it does.
Could a metaverse somehow explain why our universe exists at all?
The hope here would be that the metaverse provides a context in which explanatory language becomes appropriate, just as cause-and-effect talk becomes useful in the emergent higher-level descriptions we apply to our everyday environment.

There is a problem, however: unlike the straightforward cosmological multiverse, which can arise naturally in theories of inflationary cosmology, different elements of the metaverse are not actually connected to each other by dynamical processes or influences. 
One reality is not created from another one, so it is hard to envision a sense in which the collection of realities explains the existence or properties of any individual member.
The best we might do is to imagine a maximal metaverse, in which every reality exists, and then try to account for the specific one in which we live -- but as noted above, this program is faced with significant obstacles.
Moreover, the metaverse still faces a severe problem with explanatory regression, as we would be left trying to explain the existence of multiple realities rather than just one.

This is not to argue that such a metaverse could not (in some sense) exist. Even if such a thing is forever beyond our empirical reach, positing its existence could conceivably play an explanatory role in understanding the properties of our actual universe (though I personally am skeptical).
But as the metaverse itself has no reason to be a necessarily existent thing, and because dynamical processes within it cannot causally account for the creation of our universe (as the different elements of the metaverse are non-interacting by hypothesis), it does not directly provide an answer to the question of why there is something rather than nothing.

\begin{itemize}
\item \textit{Principle.} 
\end{itemize}

Aside from an actual being or metaverse that could account for the existence of reality, we might imagine that the best explanation takes the form of a principle that picks out our universe among all the conceivable ones.
Perhaps our universe is the simplest subject to certain conditions, or perhaps all possible realities actually exist.
Such an answer would again face the explanatory regression problem, as noted, but the existence of such a principle could arguably serve to soften the blow of the universe not being unique or necessary.
The biggest obstacle is that it's hard to see, given what we know about the actual universe, what such a principle could possibly be.
Future scientific discoveries could reveal such an answer.

\begin{itemize}
\item \textit{Coherence.} 
\end{itemize}

One option we haven't considered in detail is the idea that ``nothing exists'' might not, despite the seeming naturalness of the formulation, actually be a coherent idea.
We are once again challenged to think beyond our experience of objects within the physical world, where it is very sensible to imagine that an existent thing (say, the desk on which I am typing) could counterfactually have not existed.
It is tempting, if only by analogy, to imagine the same thing about the universe as a whole; Parfit \cite{parfit} refers to the possibility that ``nothing would have been." 

But what does ``been'' really mean in such a construction?
The absence of anything existing isn't quite the same as ``that which exists" being identified with ``nothing," as ``existing" isn't something that ``nothing" can sensibly do.
Things can fail to exist within reality, but it does not immediately follow that reality itself could have not existed.
(Could we assign a probability to that having been true?)
So perhaps the universe exists simply because there was no coherent alternative.

Suggesting this possibility requires that we look beyond the naive metaphor of a reality-chooser, among whose potential options was to choose ``nothing." 
Perhaps our language and modes of thought are tricking us, and existence is something that is metaphysically unavoidable.
In that case some form of reality would be necessary, even if the specific form were left unexplained; we would still face the challenge of understanding our actual universe.

\begin{itemize}
\item \textit{Brute fact.} 
\end{itemize}

Every attempt to answer the question ``Why is there something rather than nothing?" ultimately grounds in a brute fact, a feature of reality that has no further explanation.
The universe is not unique, and there are no necessary beings; even if we decide that the concept of nothingness is incoherent, at least some properties of our particular universe are ultimately contingent.
By the standards of modern science, it is extremely hard to see what could possibly qualify as a final and conclusive ``reason why" the universe exists.

Perhaps the absence of such a reason shouldn't be surprising.
As our knowledge of the universe improves, questions that once seemed urgent can become un-asked, as we realize that the context in which they were posed was not appropriate. 
In Kepler's time, the question of why there were precisely six planets (Mercury, Venus, Earth, Mars, Jupiter, and Saturn) was a natural one to ask, and he proposed a model where the five Platonic solids were inscribed between their orbits.
Today we know that there are more than five planets, that the definition of a ``planet'' is controversial, and that a certain amount of randomness is involved in accounting for the actual distribution of bodies in the Solar System.
In the 1930's, Eddington attempted to derive a numerical formula that would explain why the fine-structure constant $\alpha$ of electromagnetism should be exactly 1/136, as it was suspected to be at the time; when experiments improved the measured value to something closer to 1/137, he adjusted his formula in response \cite{eddington}.
Today we know that $\alpha$ is not the reciprocal of any integer.
It may still be true, of course, that there exists a subtle and elegant formula yet to be discovered that exactly reproduces the value of $\alpha$, but in modern physics where electromagnetism is subsumed into a broader context of quantum field theory and electroweak unification, the search for such a formula is not a priority.

The universe could be the same way.
Perhaps at bottom its existence and specific features include brute facts that are in some sense completely arbitrary; or perhaps there is a deeper principle that explains why it is precisely this universe, and the only brute fact is the validity of that principle.
We are always welcome to look for deeper meanings and explanations.
What we can't do is demand of the universe that there be something we humans would recognize as a satisfactory reason for its existence.

\section*{Acknowledgements}
 
This research is funded in part by the Walter Burke Institute for Theoretical Physics at Caltech and by DOE grant DE-SC0011632.

\end{document}